\documentclass[lettersize,journal]{IEEEtran}
\usepackage{amsmath,amsfonts}
\usepackage{amssymb}
\usepackage{algorithmic}
\usepackage{algorithm}
\usepackage{array}
\usepackage{color,xcolor,soul}
\usepackage{textcomp}
\usepackage{stfloats}
\usepackage{url}
\usepackage{verbatim}
\usepackage{graphicx}
\usepackage{subcaption}
\usepackage{cite}
\usepackage[absolute,showboxes]{textpos}
\usepackage{hyperref}

\TPMargin{5pt}

\newcommand{\copyrightstatement}{
    \begin{textblock}{0.84}(0.06,0.01)    
         \noindent
         \footnotesize
         \copyright  2024 IEEE.  Personal use of this material is permitted.  Permission from IEEE must be obtained for all other uses, in any current or future media, including reprinting/republishing this material for advertising or promotional purposes, creating new collective works, for resale or redistribution to servers or lists, or reuse of any copyrighted component of this work in other works. The peer-reviewed paper is available at \url{https://doi.org/10.1109/LCOMM.2024.3495998}.
    \end{textblock}
}

\newcommand{\argmin}{\mathop{\mathrm{argmin}}} 
\DeclareMathOperator{\vect}{vec}

\begin{document}
\copyrightstatement
\bstctlcite{IEEEexample:BSTcontrol}
\title{Semi-blind Channel Estimation for Massive MIMO LEO Satellite Communications}
\author{Abdollah~Masoud~Darya,~\IEEEmembership{Graduate~Student~Member,~IEEE,}
        and~Saeed~Abdallah,~\IEEEmembership{Member,~IEEE}
\thanks{Abdollah Masoud Darya is with SAASST and the Department of Electrical Engineering, University of Sharjah, Sharjah, UAE (email: abdollah.masoud@ieee.org). Saeed Abdallah is with the Department of Electrical Engineering, University of Sharjah, Sharjah, UAE (email: sabdallah@sharjah.ac.ae).}
\thanks{Manuscript received Month 00, 0000; revised Month 00, 0000.}}
%


\maketitle

\begin{abstract}
This letter proposes decision-directed semi-blind channel estimation for massive multiple-input multiple-output low-Earth-orbit satellite communications. Two semi-blind estimators are proposed. The first utilizes detected data symbols in addition to pilot symbols. The second, a modified semi-blind estimator, is specially designed to mitigate the channel-aging effect caused by the highly dynamic nature of low-Earth-orbit satellite communication channels---an issue that adversely impacts the performance of pilot-based estimators. Consequently, this modified estimator outperforms an optimal pilot-based estimator in terms of normalized mean square error and achieves symbol error rate performance comparable to that of a Genie-aided (perfectly known channel) detector. The trade-offs between the proposed estimators are also examined.
\end{abstract}


\section{Introduction}
Direct communications between satellites and mass market terrestrial user terminals (UTs), such as smartphones, are becoming a reality \cite{laursen2023no}. This so-called direct-to-cell connectivity leverages satellites as cell towers in space and is an essential component of future 6G non-terrestrial networks \cite{tuzi2023satellite}. This technology aims to address the coverage gaps of terrestrial communications to unserved and underserved communities. Several startups aim to tackle this task by utilizing low-Earth-orbit (LEO) satellites instead of geostationary orbit (GEO) satellites because of their shorter production cycle and cheaper launch costs. Compared to GEO satellites, communication through LEO satellites is characterized by lower latency but higher Doppler shifts \cite{3gpp}.\par
The satellite-motion-induced Doppler shift and delay can be compensated using data from the satellite's ephemeris in addition to the user's position derived from Global Navigation Satellite System (GNSS) receivers \cite{3gpp}. Yet, other effects, such as the Doppler shift caused by the user's movement and the propagation delays induced by the multipath effect are more difficult to compensate since they depend on the user's local environment, which---for a moving user---is continuously changing.\par
The seminal work by You \textit{et al.} \cite{you2020massive} introduced the use of multibeam transmission, massive multiple-input multiple-output (mMIMO), and orthogonal frequency-division multiplexing (OFDM) to enhance the spectral efficiency of LEO satellite communications. To limit the impact of interference introduced by these features, an accurate estimation of the channel state information (CSI) is required. Recent advancements in mMIMO LEO satellite channel estimation include pilot-based minimum mean square error (MMSE) estimators \cite{shen2022random,abdelsadek2022distributed,ying2023quasi,li2023channel}, and a data-aided variational Bayesian inference estimator \cite{wang2022joint}.\par
The user's multi-path delays and movement-induced Doppler shifts cause \emph{channel aging}, whereby the CSI obtained through pilot-based estimation becomes outdated with time \cite{chopra2017performance}. One solution to this issue is to send pilot symbols more frequently to obtain updated channel estimates. Yet, this solution adds considerable overhead to the system.\par
Other methods have been proposed in the literature to address the channel aging problem in mMIMO LEO satellite communications. For instance, Yue \textit{et al.} \cite{yue2022block} proposed Kalman channel tracking to compensate for the aging effect. They proposed a block-based Kalman tracking scheme that outperformed the more complex symbol-based scheme for high signal-to-noise ratio (SNR). Yet, Kalman channel tracking requires prior knowledge of the channel and noise statistics \cite{zhu2007channel}. Alternatively, Zhang \textit{et al.} \cite{zhang2021deep} proposed a deep-learning-based scheme to compensate for the channel aging effect. Their proposed scheme is based on the long short-term memory (LSTM) architecture. Their model was trained to capture the correlation of channel variations, so that it may take the current CSI as an input and predict future channels. Yet, deep learning methods are sensitive to the quality of training data and demand extensive offline training \cite{hu2020deep}.\par
This work proposes decision-directed semi-blind (DD-SB) channel estimation as a method of obtaining updated channel estimates without transmitting additional pilot symbols. The proposed method does not assume prior knowledge of the channel and noise statistics and does not require offline training. In semi-blind channel estimation, the pilot-based channel estimate is used to detect data symbols. These data symbols are then used, in addition to the pilot symbols, to acquire an updated channel estimate. Additionally, a modified DD-SB (MDD-SB) estimator is proposed that minimizes the complexity of the DD-SB estimator and mitigates the effects of channel aging by periodically updating the channel estimate using only the most recent detected data symbols. Comparisons between the proposed MDD-SB estimator, an optimal pilot-based estimator, and a Genie-aided detector are performed in terms of the normalized mean square error (NMSE) and symbol error rate (SER). The complexity of the proposed estimators is also provided.\par
The rest of this letter is organized as follows: Section \ref{ChannelModel} presents the system model, Section \ref{ChannelEstimation} investigates the proposed semi-blind channel estimators, Section \ref{RnD} discusses the results, and Section \ref{Conc} concludes the letter.\par


\section{System Model}\label{ChannelModel}
This work considers a mMIMO LEO satellite communications system serving $K$ single-antenna user terminals (UTs) simultaneously. The LEO satellite is equipped with a uniform planar array (UPA) consisting of $M=M_xM_y$ elements, where the numbers of $x$- and $y$-axis antenna elements are represented by $M_x$ and $M_y$, respectively. The UPA supports multi-beam transmission with full frequency reuse. The system operates in the frequency division duplexing mode and employs OFDM \cite{you2020massive}.\par
The uplink channel between the satellite and user terminal (UT) $k$, at time $t$ and frequency $f$, consists of two components: line of sight (LoS) and non-line of sight (NLoS) \cite{yue2022block}, i.e,
\begin{equation}\label{eq1}
\boldsymbol{h}_k(t,f)=\sqrt{\beta_k}\left(h_k^{\text{LoS}}(t,f)+h_k^{\text{NLoS}}(t,f)\right)\cdot\boldsymbol{a}_{k}.
\end{equation}
The path loss component ${\beta_k}$ is represented by \cite{3gpp}
\begin{equation}
    {\beta_k}=32.45+20\log_{10}(f_c)+20\log_{10}(d_k),
\end{equation}
where $f_c$ is the carrier frequency, and $d_k$ is the distance between the satellite and UT $k$. Furthermore, $\boldsymbol{a}_{k}$ is the UPA response vector represented by \cite{you2020massive}
\begin{equation}
\boldsymbol{a}_{k} = \boldsymbol{v}_x\left(\sin\left(\theta^y_k\right)\cos\left(\theta^x_k\right)\right) \otimes \boldsymbol{v}_y\left(\cos\left(\theta^y_k\right)\right) \in \mathbb{C}^{M\times 1}~,
\end{equation}
where $\theta^y_k$ and $\theta^x_k$ are the angles associated with the $x$- and $y$-axis of the propagation paths for UT $k$ (due to the satellite’s high altitude, all paths for a given UT share the same angles \cite{you2020massive}), respectively, and $\otimes$ represents the Kronecker product. The array vectors $\boldsymbol{v}_x$ and $\boldsymbol{v}_y$, where $d\in\left\{x,y\right\}$ and $\boldsymbol{v}_d\left(\mathcal{D}\right) \in \mathbb{C}^{M_d\times 1}$, can be represented as
\begin{equation}
\boldsymbol{v}_d\left(\mathcal{D}\right) = \frac{\left[1, \exp\left\{-j\pi\mathcal{D}\right\}, \cdots, \exp\left\{-j\pi\left(M_d-1\right)\mathcal{D}\right\}\right]^T}{\sqrt{M_d}}.
\end{equation}
The vector $\boldsymbol{a}_{k}$ can be deduced from known UT and satellite positions \cite{3gpp}, or estimated using the methods in \cite{boukhedimi2023angle} if the positions are unknown.\par
The LoS channel component is represented by \cite{zhang2022deep}
\begin{equation}
\begin{split}    
h_k^{\text{LoS}}(t,f)=\sqrt{\frac{\kappa_k}{\kappa_k+1}}\cdot\exp\left\{-j2\pi f\tau_k^{\text{LoS}}\right\}\\\cdot\exp\left\{j2\pi t\left(\nu_k^{\text{SAT-LoS}}+\nu_k^{\text{UT-LoS}}\right)\right\},
\end{split}
\end{equation}
where $\kappa_k$ is the Rician factor, $\tau_k^{\text{LoS}}$ is the LoS time delay, $\nu_k^{\text{SAT-LoS}}$ is the LoS satellite (SAT) Doppler shift, and $\nu_k^{\text{UT-LoS}}$ is the LoS user Doppler shift. Due to the high velocity and altitude of the LEO satellite $\nu_k^{\text{SAT-LoS}}\gg\nu_k^{\text{UT-LoS}}$.\par
The NLoS channel component is represented by \cite{zhang2022deep}
\begin{equation}
\begin{split}
h_k^{\text{NLoS}}(t,f)=&\sqrt{\frac{1}{\kappa_k+1}}\cdot\sqrt{\frac{1}{P_k}}\sum_{p=1}^{P_k}g_{k,p}\cdot\exp\left\{-j2\pi f\tau_{k,p}^{\text{NLoS}}\right\}\\&\cdot\exp\left\{j2\pi t\left(\nu_{k,p}^{\text{SAT-NLoS}}+\nu_{k,p}^{\text{UT-NLoS}}\right)\right\}.
\end{split}
\end{equation}
Note that $g_{k,p}$ is the Rayleigh fading complex-valued gain with real and imaginary parts that are independently and identically Gaussian-distributed with zero mean and unit variance \cite{abdelsadek2022distributed}. Furthermore, $P_k$ is the number of NLoS propagation paths per UT $k$, $\tau_{k,p}^{\text{NLoS}}$ is the NLoS time delay per user $k$ and path $p$, and $\nu_{k,p}^{\text{SAT-NLoS}}$ and $\nu_{k,p}^{\text{UT-NLoS}}$ are the NLoS satellite and user Doppler shifts, respectively. Due to the high altitude of the satellite, the satellite's Doppler component is identical for all paths, therefore $\nu_{k,p}^{\text{SAT-NLoS}} = \nu_k^{\text{SAT-LoS}} = \nu_k^{\text{SAT}}$ \cite{you2020massive}.\par
By simplifying the previous expressions, we obtain
\begin{equation}
\begin{split} 
\boldsymbol{h}_k(t,f)=&\sqrt{\frac{\beta_k}{\kappa_k+1}}\cdot\exp\left\{j2\pi t\nu_{k}^{\text{SAT}}\right\}\\&\cdot\left(h_k^{\text{LoS}}(t,f)+h_k^{\text{NLoS}}(t,f)\right)\cdot\boldsymbol{a}_{k},
\end{split} 
\end{equation}
where
\begin{equation}
h_k^{\text{LoS}}(t,f)=\sqrt{\kappa_k}\cdot\exp\left\{j2\pi \left(t\nu_k^{\text{UT-LoS}}-f\tau_k^{\text{LoS}}\right)\right\},
\end{equation}
and
\begin{equation}
h_k^{\text{NLoS}}(t,f)\!=\!\sqrt{\frac{1}{P_k}}\sum_{p=1}^{P_k}g_{k,p}\cdot\exp\left\{j2\pi\!\left(t\nu_{k,p}^{\text{UT-NLoS}}\!-\!f\tau_{k,p}^{\text{NLoS}}\right)\!\right\}\!.
\end{equation}
Note that $\tau_{k,p}^{\text{NLoS}}=\tau_k^{\text{LoS}}+\tau_{k,p}^{\text{MP}}$, where $\tau_{k,p}^{\text{MP}}$ is the multipath (MP) time delay per path $k$. Furthermore, $\tau_k^{\text{LoS}}\gg\tau_{k,p}^{\text{MP}}$.\par
The demodulated uplink received signal over subcarrier $c$ of OFDM symbol $s$ is represented by \cite{you2020massive} 
\begin{equation}
\label{eq6}
\boldsymbol{y}_{s,c}= \sum_{k=1}^{K}\boldsymbol{h}_{k,s,c}\cdot x_{k,s,c}+\boldsymbol{z}_{s,c} \in\mathbb{C}^{M\times 1}~,
\end{equation}
where $\boldsymbol{h}_{k,s,c}=h_{k}\left(s\left(T_{sl}+T_{cp}\right),c/T_{sl}\right)\cdot\boldsymbol{a}_k$, $x_{k,s,c} \in \mathbb{C}$ is the symbol transmitted by UT $k$, and $\boldsymbol{z}_{s,c} \sim \mathcal{C}\mathcal{N}\left(\boldsymbol{0},\sigma^2\boldsymbol{I}\right)$ is the additive white Gaussian noise (AWGN) with variance $\sigma^2$. Note that the symbol duration $T_{sl}=N_{sc}T_s$, where $N_{sc}$ is the number of subcarriers and $T_s$ is the sampling period. Furthermore, the cyclic prefix duration is represented by $T_{cp}=N_{cp}T_s$, where $N_{cp}$ is the cyclic prefix length.\par

All observations at the satellite side for $S$ symbols for all $K$ UTs are collected into the matrix $\boldsymbol{Y}_c \in \mathbb{C}^{S\times M}$, where $\boldsymbol{Y}_c=\left[\boldsymbol{y}_{1,c},\dots,\boldsymbol{y}_{S,c}\right]^T$. Additionally 
\begin{equation}
    \boldsymbol{Y}_c\!=\!\left(\boldsymbol{H}_c \odot \left(\boldsymbol{X}_c \odot \exp\left\{-j2\pi \boldsymbol{T}\otimes\boldsymbol{\mathcal{V}}_c^\text{SAT}\right\}\right)\right)^T\!\cdot\!\boldsymbol{A}\!+\!\boldsymbol{Z}_c,
\end{equation}
where $\odot$ denotes the Hadamard product. Furthermore,
\begin{equation}
\begin{split}
    \boldsymbol{H}_c=& \begin{bmatrix} 
    h_{1,1,c} & \dots & h_{1,S,c}\\
    \vdots & \ddots & \vdots\\
    h_{K,1,c} & \dots & h_{K,S,c}
    \end{bmatrix}  \in \mathbb{C}^{K\times S},\\
    \boldsymbol{X}_c=& \begin{bmatrix}
    x_{1,1,c} & \dots & x_{1,S,c}\\
    \vdots & \ddots & \vdots\\
    x_{K,1,c} & \dots & x_{K,S,c}
    \end{bmatrix} \in \mathbb{C}^{K\times S},\\
    \boldsymbol{A}=& \begin{bmatrix}
    \boldsymbol{a}_{1}\\
    \vdots\\
    \boldsymbol{a}_{K}
    \end{bmatrix} \in \mathbb{C}^{K\times M},
\end{split}
\end{equation}
and $\boldsymbol{T}=\left[T_{sl}+T_{cp},\dots,S\left(T_{sl}+T_{cp}\right)\right]$. Furthermore, $\boldsymbol{\mathcal{V}}_c^{\text{SAT}}=\left[\nu_{1}^{\text{SAT}},\dots,\nu_{K}^{\text{SAT}}\right]^T$ represents the satellite Doppler compensation performed at the UT side \cite{you2020massive}, and $\boldsymbol{Z}_c\in \mathbb{C}^{S\times M}$ is the AWGN matrix.\par

\section{Proposed Channel Estimators}\label{ChannelEstimation}
In the uplink frame, each user transmits $P$ orthogonal pilot symbols based on Zadoff-Chu sequences \cite{li2023channel} to the satellite in a block-type pilot arrangement \cite{coleri2002channel}, followed by $D$ data symbols. The pilot symbols are used to conduct pilot-based channel estimation, while the data symbols in addition to the pilots are used to conduct DD-SB channel estimation. Using the uplink estimate from this work, the downlink estimate can be found through the deep learning method proposed in \cite{zhang2022deep}.\par
\subsection{Pilot-based Estimator} 
The pilot-based least-squares (P-LS) estimator is given as
\begin{equation}
\boldsymbol{\hat{H}}{}_c^\text{P}=\argmin_{\boldsymbol{\hat{H}}{}_c}~\left\Vert\boldsymbol{Y}{}_c^\text{P}-\left(\boldsymbol{\hat{H}}{}_c\odot\boldsymbol{X}_c^\text{P}\right)\boldsymbol{A}\right\Vert^2.
\end{equation}
Therefore,
\begin{equation}
\boldsymbol{\hat{H}}{}_c^\text{P}=\left(\boldsymbol{Y}{}_c^\text{P}\boldsymbol{A}^{+}\right)^T\oslash\boldsymbol{X}_c^\text{P},
\end{equation}
where $(\cdot)^{+}$ is the Moore-Penrose (pseudo) inverse, $\oslash$ denotes the Hadamard division, and $\boldsymbol{Y}{}_c^\text{P}$ is the received signal over pilot symbols $\boldsymbol{X}_c^\text{P}$.\par
The mean P-LS channel estimate across all pilot symbols $P$ is considered, and then duplicated by the number of transmitted data symbols $D$, i.e., $\boldsymbol{\hat{\bar{H}}}{}_c^\text{P}\in\mathbb{C}^{K\times D}$. This means that the estimator considers the channel as time-invariant for $P$ symbols. This step is performed for two main reasons. First, it is used to address the potential mismatch between the number of pilot and data symbols during the data detection stage. Second, it reduces the impact of AWGN on the channel estimate for low SNR.\par
\subsection{Decision-Directed Semi-blind Estimator}\label{SB-LS}
Semi-blind estimators offer two main advantages over pilot-based estimators. First, semi-blind estimators can be used to obtain more accurate channel estimates as they utilize the pilots in addition to the detected data symbols \cite{nayebi2017semi}. Second, pilot-based estimates will become outdated with time due to channel aging. Therefore, a DD-SB estimator that utilizes up-to-date detected data symbols in addition to pilots can provide better estimation performance in the long-term.\par
The P-LS channel estimate $\boldsymbol{\hat{\bar{H}}}{}_c^\text{P}$ is used to perform zero-forcing (ZF) equalization on the received data symbols, where
\begin{equation}
\boldsymbol{\hat{X}}{}_c^\text{D}=\argmin_{\boldsymbol{\hat{X}}{}_c}~\left\Vert\boldsymbol{Y}{}_c^\text{D}-\left(\boldsymbol{\hat{\bar{H}}}{}_c^\text{P}\odot\boldsymbol{\hat{X}}{}_c\right)\boldsymbol{A}\right\Vert^2.
\end{equation}
Therefore,
\begin{equation}\label{eq_ref1}
\boldsymbol{\hat{X}}{}_c^\text{D}=\left(\boldsymbol{Y}{}_c^\text{D}\boldsymbol{A}^{+}\right)^T\oslash\boldsymbol{\hat{\bar{H}}}{}_c^\text{P},
\end{equation}
where $\boldsymbol{Y}{}_c^\text{D}$ is the received signal due to the transmission of data symbols $\boldsymbol{X}_c^\text{D}$. Next, the data symbols are detected using a minimum distance detector \cite{zakharov2009optimal} and collected into $\boldsymbol{\hat{\bar{X}}}{}_c^\text{D}$. In turn, $\boldsymbol{\hat{\bar{X}}}{}_c^\text{D}$ is used in addition to the known pilot symbols $\boldsymbol{X}{}_c^\text{P}$ to find the DD-SB channel estimate
\begin{equation}
\boldsymbol{\hat{H}}{}_c^\text{SB}=\left(
\begin{bmatrix}\boldsymbol{Y}{}_c^\text{P}\\\boldsymbol{Y}{}_c^\text{D}\end{bmatrix}\boldsymbol{A}^{+}\right)^T \oslash\left[\boldsymbol{X}{}_c^\text{P},\boldsymbol{\hat{\bar{X}}}{}_c^\text{D}\right].
\end{equation}
Similar to the P-LS estimate, the mean channel estimate $\boldsymbol{\hat{\bar{H}}}{}_c^\text{SB}$ across all transmitted pilot and data symbols is considered.\par
It is important to note that $\boldsymbol{A}^{+}$ is the most complex operation of the P-LS and DD-SB channel estimators, with complexity in the order of $\mathcal{O}\left(K^2M\right)$. Since $\boldsymbol{A}$ is assumed to be time-invariant, this operation can be performed offline once the position of the satellite or UT changes considerably \cite{you2020massive}. The run-time complexity of the P-LS estimator per subcarrier is $\mathcal{O}\left(PMK+PK\right)$, while the complexity of the DD-SB detection step per subcarrier is $\mathcal{O}\left(DMK+DK\right)$, and estimation step is $\mathcal{O}\left((P+D)MK+(P+D)K\right)$.\par
With channel aging in effect, the initial P-LS estimate deteriorates with time. This, as a result, leads to the decreased accuracy of the DD-SB estimator since it includes the observed pilot signal in its formulation. This motivated the development of the estimator presented in the following subsection.\par

\begin{figure}
\centering
\begin{subfigure}[b]{1\linewidth}
   \caption{}
   \includegraphics[width=\linewidth]{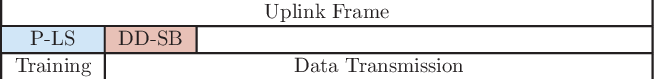}
\end{subfigure}
\begin{subfigure}[b]{1\linewidth}
   \caption{}
   \includegraphics[width=\linewidth]{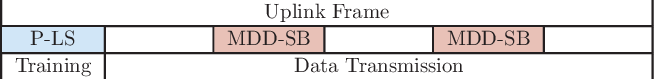}
\end{subfigure}
\caption{(a) DD-SB frame division, (b) MDD-SB frame division.}\label{Fig_a}
\end{figure}

\subsection{Modified Decision-Directed Channel Estimator}
The DD-SB channel estimator introduced in Section \ref{SB-LS} can be optimized to improve its resilience against channel aging and reduce its complexity. This is achieved by removing the outdated pilot symbols from the formulation and instead relying purely on the most recent detected data symbols to obtain up-to-date channel estimates\footnote{There is a trade-off in including detected data symbols from neighboring blocks: while this can improve estimation performance, especially in low SNR scenarios, using older symbols risks amplifying the channel aging effect, particularly in high SNR scenarios.}. This method is referred to henceforth as the MDD-SB estimator.\par
Let the uplink frame be divided into $N+1$ blocks, where block $\imath=0$ corresponds to the pilot transmission and channel estimation phase, and blocks $\imath=[1,\cdots,N]$ correspond to the data reception, detection, and semi-blind channel estimation phase. The P-LS channel estimate, generated at block $0$, is used along with the detected data symbols for $\imath=1$ to perform DD-SB channel estimation at the end of $\imath=1$. In contrast, MDD-SB channel estimation can be performed at some or all $\imath>0$ blocks, as it relies solely on the detected data symbols. Fig. \ref{Fig_a} illustrates the difference between the DD-SB and the MDD-SB methods in terms of the uplink frame division.\par
Performing MDD-SB channel estimation for each $\imath>0$ block will yield the best estimation performance while imposing the highest computational overhead. Selecting the best interval for MDD-SB estimation depends on the Doppler spread, satellite computational capabilities, and estimation accuracy requirements. The optimal choice of interval for a particular system can be found offline before implementation. The run-time complexity of the detection and estimation steps per subcarrier for a single MDD-SB iteration, which is $\mathcal{O}\left(DMK+DK\right)$, is linear in terms of $D,M,$ and $K$. If applied frequently, the MDD-SB estimator's timing overhead may eventually surpass that of the DD-SB estimator, though this increase is minor and outweighed by improved estimation accuracy.\par
The proposed MDD-SB method is outlined in Algorithm \ref{alg1}, for blocks $\imath=[n,\cdots,N]$. It utilizes the initial P-LS channel estimate $\boldsymbol{\hat{\bar{H}}}{}_c^\text{P}$ to detect the received data symbols $\boldsymbol{\hat{\bar{X}}}{}_c^{\text{D},\imath}$ in block $\imath$. These detected data symbols are then used to update the channel estimate $\boldsymbol{\hat{\bar{H}}}{}_c^\text{new}$, which is then used to detect the received data symbols in the next block $\boldsymbol{\hat{\bar{X}}}{}_c^{\text{D},\imath+1}$. This process is repeated to continuously update the channel estimate and mitigate the channel aging effect. The following section presents the performance results of the proposed method, compared against an optimal pilot-based estimator and a Genie-aided detector.\par

\begin{algorithm}[!t]
\caption{Proposed MDD-SB Channel Estimation}
\begin{algorithmic}[1]
\renewcommand{\algorithmicrequire}{\textbf{Input:}}
\renewcommand{\algorithmicensure}{\textbf{Output:}}
\REQUIRE Pilot-based channel estimate $\boldsymbol{\hat{\bar{H}}}{}_c^\text{P}$
\ENSURE Updated channel estimate $\boldsymbol{\hat{\bar{H}}}{}_c^\text{new}$
\STATE Let $\boldsymbol{\hat{\bar{H}}}{}_c^\text{new}=\boldsymbol{\hat{\bar{H}}}{}_c^\text{P}$
\FOR {$\imath = n$ to $N$}
\STATE $\boldsymbol{\hat{X}}{}_c^{\text{D},\imath}=\left(\boldsymbol{Y}{}_c^{\text{D},\imath}\boldsymbol{A}^{+}\right)^T\oslash\boldsymbol{\hat{\bar{H}}}{}_c^\text{new}$
\STATE Detect $\boldsymbol{\hat{X}}{}_c^{\text{D},\imath} \Rightarrow \boldsymbol{\hat{\bar{X}}}{}_c^{\text{D},\imath}$
\STATE $\boldsymbol{\hat{H}}{}_c^\text{new}=\left(\boldsymbol{Y}{}_c^{\text{D},\imath}\boldsymbol{A}^{+}\right)^T \oslash\boldsymbol{\hat{\bar{X}}}{}_c^{\text{D},\imath}$
\STATE Average $\boldsymbol{\hat{H}}{}_c^\text{new} \Rightarrow \boldsymbol{\hat{\bar{H}}}{}_c^\text{new}$
\ENDFOR
\end{algorithmic}
\label{alg1}
\end{algorithm}

\section{Results and Discussion}\label{RnD}
This section presents simulation results of the proposed P-LS, DD-SB, and MDD-SB estimators in terms of NMSE and SER over $10^5$ channels. Furthermore, the MDD-SB estimator is compared with an optimal pilot-based estimator (perfect channel knowledge at block 0) in terms of NMSE, and a Genie-aided detector (perfect channel knowledge at all blocks) in terms of SER.\par
The simulation parameters utilized in this work are as follows. The satellite, equipped with a UPA consisting of $M_x=M_y=10$ antenna elements spaced at half-wavelength in the $x$- and $y$-axis, is assumed to be orbiting at $600\,\mathrm{km}$. It receives signals from $10$ UTs transmitting at a carrier frequency $f_c$ of $30\,\mathrm{GHz}$ and using 16-QAM modulation. The channel Rician factor is set to $\kappa_k=10\,\mathrm{dB}$, and the maximum number of NLoS propagation paths $P_k$ is $4$. The satellite Doppler shift is bounded by $\left\vert\nu_k^{\text{SAT}}\right\vert\leq 788\,\mathrm{kHz}$ while the LoS UT Doppler shift is bounded by $\left\vert\nu_k^{\text{UT-LoS}}\right\vert\leq 200\,\mathrm{Hz}$.\par

The reference time-varying channel $\boldsymbol{h}_{s}^\text{ref}=[h_{1,s}\cdot\exp\left\{-j2\pi t\nu_{1}^{\text{SAT}}\right\},\dots,h_{K,s}\cdot\exp\left\{-j2\pi t\nu_{K}^{\text{SAT}}\right\}]^T\in\mathbb{C}^{K\times 1}$ is used to calculate the NMSE of the channel estimate using
\begin{equation}
\text{NMSE}^\text{X}=\frac{\left\Vert\vect\left(\left[\boldsymbol{h}_{s}^\text{ref},\dots,\boldsymbol{h}_{S}^\text{ref}\right]\right)-\vect\left(\boldsymbol{\hat{\bar{H}}}{}^\text{X}\right)\right\Vert^2}{\left\Vert\vect\left(\left[\boldsymbol{h}_{s}^\text{ref},\dots,\boldsymbol{h}_{S}^\text{ref}\right]\right)\right\Vert^2},
\end{equation}
where $\text{X}\in\left\{\text{P},\text{SB}\right\}$, and depending on the method, $S$ may consist of $P$ pilot symbols, $D$ data symbols, or both ($P+D$). Furthermore, $\vect(\cdot)$ represents the vectorization operation, and the SNR is represented as
\begin{equation}
\text{SNR}=10\log_{10}\frac{\left\Vert\vect\left(\left(\left[\boldsymbol{h}_{1}^\text{ref},\dots,\boldsymbol{h}_{S}^\text{ref}\right] \odot\boldsymbol{X}\right)^T\cdot\boldsymbol{A}\right)\right\Vert^2}{\sigma^2}.
\end{equation}\par

\begin{figure}[!t]
\centering
\includegraphics[width=\linewidth]{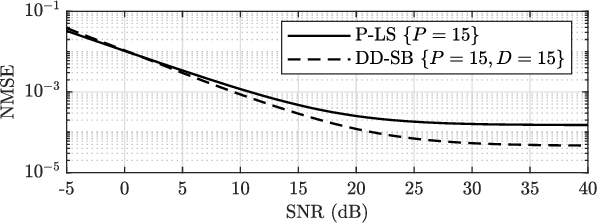}
\caption{The NMSE vs SNR trend of the P-LS and DD-SB estimators, using $P$ pilot and $D$ data symbols.}
\label{Fig_1}
\end{figure}

Fig. \ref{Fig_1} shows the NMSE performance of the DD-SB estimator (presented in Section \ref{SB-LS}) at block $\imath=1$, using $15$ pilot symbols and $15$ detected data symbols, compared to a P-LS estimator utilizing $15$ pilot symbols. The DD-SB estimator outperforms the P-LS estimator from SNR $>1\,\mathrm{dB}$, while the P-LS estimator has better performance at lower SNR. This is due to the influence of AWGN on the detected data symbols. Furthermore, both the P-LS and DD-SB trends converge to a minimum at high SNR due to the averaging step. This figure shows that the DD-SB estimator can be used to acquire more accurate channel estimates compared to a P-LS estimator by leveraging detected data symbols in addition to pilot symbols. Performance at low SNR can be enhanced by using more detected data symbols or replacing \eqref{eq_ref1} with an MMSE receiver. However, MMSE detection and estimation rely on noise statistics which are assumed to be unknown in this work.\par

\begin{figure}[!t]
\centering
\includegraphics[width=\linewidth]{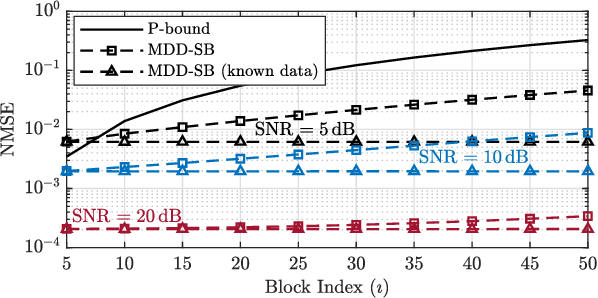}
\caption{The NMSE versus block index of the proposed MDD-SB estimator and a benchmark optimal pilot-based estimator (P-bound).}
\label{Fig_2}
\end{figure}

In Fig. \ref{Fig_2} the NMSE of the proposed MDD-SB estimator outlined in Algorithm \ref{alg1} is presented for blocks $\imath=[5,10,\cdots,50]$, where $\boldsymbol{\hat{\bar{H}}}{}_c^\text{new}$ is updated once every 5 blocks. The NMSE of the MDD-SB estimator is compared to that of an optimal pilot-based estimator (P-bound), with perfect channel knowledge at the pilot training phase (block $0$), and a benchmark MDD-SB estimator with perfect knowledge of the transmitted data symbols. The P-LS estimate is considered at block $0$ with $15$ pilot symbols, and the MDD-SB estimate at subsequent blocks, i.e., at $\imath=[5,10,\dots,50]$ using the most recent $15$ data symbols. As the MDD-SB estimator utilizes the detected data of each fifth block to update the channel estimate, it maintains a lower NMSE level compared to the P-bound. It also achieves NMSE performance that is comparable to that of an estimator with known data symbols for high SNR. While some error accumulation can be seen, the MDD-SB estimator consistently outperforms the P-bound over time. This result demonstrates that the proposed MDD-SB estimator is preferable to pilot-based estimators that necessitate more frequent pilot transmissions, as it effectively mitigates channel aging.\par

\begin{figure}[!t]
\centering
\includegraphics[width=\linewidth]{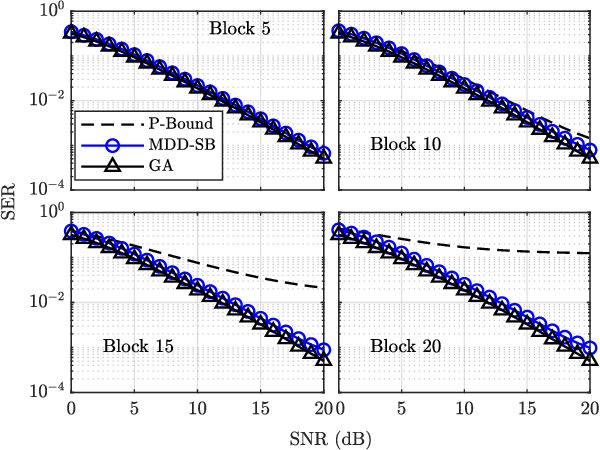}
\caption{The SER versus SNR trends of the proposed MDD-SB, P-Bound, and Genie-aided (GA) detector for blocks $\imath=[5,10,15,20]$.}
\label{Fig_3}
\end{figure}

Fig. \ref{Fig_3} compares the SER performance of the proposed MDD-SB to the P-bound, where the exact channel is perfectly known at block $0$, and a Genie-aided (GA) detector with perfect channel knowledge at every block. This plot shows the performance of each method in detecting the received data symbols at blocks $\imath=[5,10,15,20]$ with channel aging in effect. The MDD-SB estimator performs comparably to the GA detector, whereas the P-bound suffers an increase in SER with each successive block. In fact, due to channel aging, the pilot-based SER exceeds $10^{-1}$ by block $20$, thus necessitating the re-transmission of pilot symbols, while the MDD-SB estimator maintains near-optimal SER, even though the estimation was performed once every $5$ blocks. These results further highlight the potential of the proposed MDD-SB in compensating for channel aging and its superiority over pilot-based methods.\par

\section{Conclusion}\label{Conc}
This letter proposed semi-blind channel estimation for mMIMO LEO satellite communications. The DD-SB estimator outperformed the P-LS estimator in terms of NMSE, except at low SNR. The MDD-SB estimator was proposed to mitigate channel aging while reducing the complexity of the DD-SB estimator. With channel aging in effect, the MDD-SB estimator outperformed an optimal pilot-based estimator in terms of NMSE and SER, achieving SER performance comparable to a GA detector. These results indicate that the proposed MDD-SB estimator effectively mitigates channel aging by periodically updating the channel estimate, thus alleviating the need for frequent pilot transmissions necessitated by pilot-based estimators.\par

\bibliographystyle{IEEEtran}
\bibliography{Manuscript.bib}

\end{document}